\PassOptionsToPackage{table}{xcolor}
\documentclass[floatfix,lengthcheck,showpacs,amssymb,amsmath,amsfonts,twocolumn]{aastex62}

\usepackage{lineno}
\usepackage[utf8]{inputenc}
\usepackage{color}
\usepackage{graphicx}
\usepackage{float}
\usepackage{subfigure}
\usepackage{amsmath}
\usepackage{acronym}
\usepackage{multirow}
\usepackage{tikz}
\usetikzlibrary{arrows,shapes,trees,decorations.pathreplacing}
\usepackage{import}
\usepackage{svn-multi}
\usepackage{lineno}
\usepackage{hyperref}
\usepackage{gensymb}
\hypersetup{
    colorlinks=true,
    linkcolor=blue,
    filecolor=magenta,
    urlcolor=cyan,
}

\newcommand{\chieff}{\ensuremath{\chi_{\mathrm{eff}}}}
\newcommand{\rankingstat}{\ensuremath{\tilde{\Lambda}}}

\newcommand{\msun}{\ensuremath{\mathrm{M}_{\odot}}}
\newcommand{\pastro}{\ensuremath{p_{\mathrm{astro}}}}
\newcommand{\release}{\texttt{\url{www.github.com/gwastro/2-ogc}}}

\newcommand{\ias}{\hyperref[prev]{\textsuperscript{z}}}
\newcommand{\gwtc}{\hyperref[prev]{\textsuperscript{x}}}
\newcommand{\ogc}{\hyperref[prev]{\textsuperscript{y}}}

\begin{document}
\title[]{2-OGC: Open Gravitational-wave Catalog of binary mergers from analysis of public Advanced LIGO and Virgo data}

\correspondingauthor{Alexander H. Nitz}
\email{alex.nitz@aei.mpg.de}

\author[0000-0002-1850-4587]{Alexander H. Nitz}
\affil{Max-Planck-Institut f{\"u}r Gravitationsphysik (Albert-Einstein-Institut), D-30167 Hannover, Germany}
\affil{Leibniz Universit{\"a}t Hannover, D-30167 Hannover, Germany}

\author[0000-0003-1354-7809]{Thomas Dent}
\author[0000-0002-4289-3439]{Gareth S. Davies}
\affil{Instituto Galego de F\'{i}sica de Altas Enerx\'{i}as, Universidade de Santiago de Compostela, 15782 Santiago de Compostela, Galicia, Spain }

\author[0000-0002-6404-0517]{Sumit Kumar}
\author[0000-0002-0355-5998]{Collin D. Capano}
\affil{Max-Planck-Institut f{\"u}r Gravitationsphysik (Albert-Einstein-Institut), D-30167 Hannover, Germany}
\affil{Leibniz Universit{\"a}t Hannover, D-30167 Hannover, Germany}

\author[0000-0002-5304-9372]{Ian Harry}
\affil{University of Portsmouth, Portsmouth, PO1 3FX, United Kingdom}
\affil{Kavli Institute of Theoretical Physics, UC Santa Barbara, CA}
\author[0000-0002-8855-2509]{Simone Mozzon}
\affil{University of Portsmouth, Portsmouth, PO1 3FX, United Kingdom}
\author[0000-0001-7472-0201]{Laura Nuttall}
\affil{University of Portsmouth, Portsmouth, PO1 3FX, United Kingdom}
\author[0000-0002-0363-4469]{Andrew Lundgren}
\affil{University of Portsmouth, Portsmouth, PO1 3FX, United Kingdom}

\author[0000-0002-5354-5683]{M\'{a}rton T\'{a}pai}
\affil{Department of Experimental Physics, University of Szeged,
Szeged, 6720 D\'{o}m t\'{e}r  9., Hungary}


\begin{abstract} 
We present the second Open Gravitational-wave Catalog (2-OGC) of compact-binary coalescences, obtained from the complete set of public data from Advanced LIGO's first and second observing runs. For the first time we also search public data from the Virgo observatory.
The sensitivity of our search benefits from updated methods of ranking candidate events including the effects of non-stationary detector noise and varying network sensitivity; in a separate targeted binary black hole merger search we also impose a prior distribution of binary component masses.
We identify a population of 14 binary black hole merger events with probability of astrophysical origin $> 0.5$ as well as the binary neutron star merger GW170817.
We confirm the previously reported events GW170121, GW170304, and GW170727 and also report GW151205, a new marginal binary black hole merger with a primary mass of $67^{+28}_{-17}\,\msun$ that may have formed through hierarchical merger. We find no additional significant binary neutron star merger or neutron star--black hole merger events. To enable deeper follow-up as our understanding of the underlying populations evolves, we make available our comprehensive catalog of events, including the sub-threshold population of candidates and posterior samples from parameter inference of the 30 most significant binary black hole candidates.
\end{abstract}

\section{Introduction}
The Advanced LIGO~\citep{TheLIGOScientific:2014jea} and Virgo~\citep{TheVirgo:2014hva} observatories have ushered in the age of gravitational-wave astronomy. The first and second observing runs (O1 and O2) of Advanced LIGO and Virgo covered the period from 2015-2017. This provided a total of 171 days of multi-detector observing time. To date, these instruments have observed a population of binary black holes and a single binary neutron star, GW170817, which has become one of the most observed astronomical events~\citep{GBM:2017lvd}. Ten binary black hole mergers and a single binary neutron star merger have been reported in this period by the LIGO and Virgo Collaborations~\citep{LIGOScientific:2018mvr}. Several independent analyses have examined publicly released data~\citep{Nitz:2018imz,Antelis:2018smo,Venumadhav:2019tad}, including an analysis targeting binary black hole mergers that reported several additional candidates~\citep{Venumadhav:2019lyq}.
\begin{figure*}[t]
  \centering
    \includegraphics[width=2.2\columnwidth]{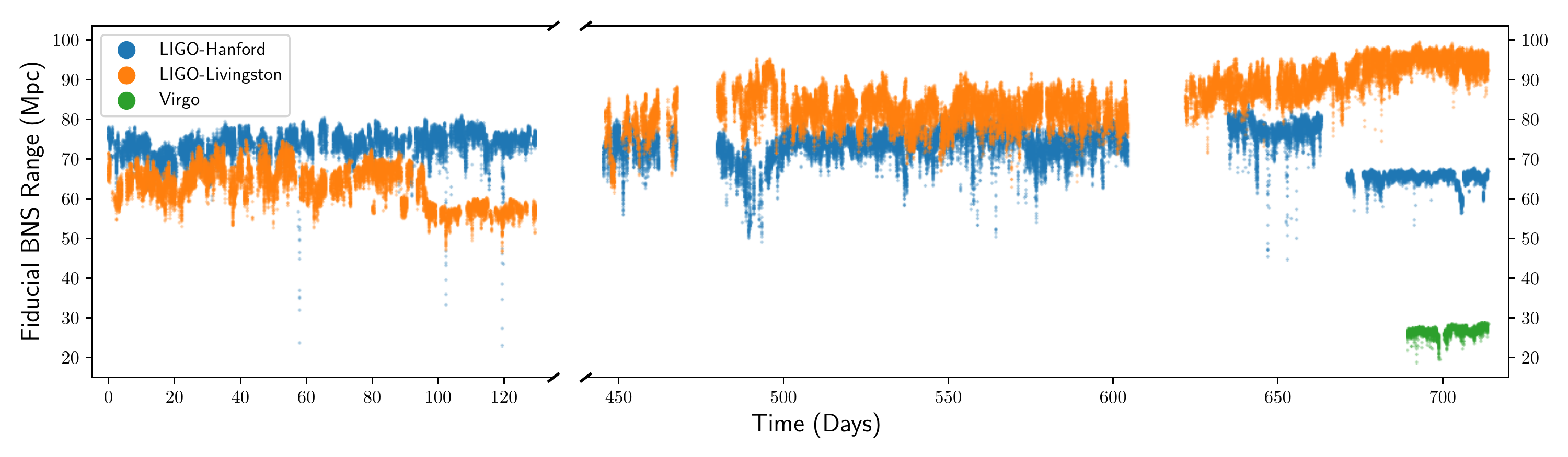}
\caption{The sky- and orientation-averaged distance up to which each observatory can detect a 1.4-1.4\,\msun BNS merger with expected SNR equal to $8$ over the O1 and O2 observing runs. The Virgo observatory did not participate in the first observing run, but joined towards the end of the second observing run.}
\label{fig:range}
\end{figure*}

The first open gravitational-wave catalog (1-OGC) searched for compact-binary coalescences during O1~\citep{Nitz:2018imz}. We extend that analysis to cover both O1 and O2 while incorporating Virgo data for the first time. During the first observing run, only the two LIGO instruments were observing. Joint three-detector observing with the Virgo instrument began in August 2017 during the second observing run.

We make additional improvements to our search by accounting for short-time variations in the network sensitivity and power spectral density (PSD) estimates directly in our ranking of candidate events. A similar procedure for tracking PSD variations was independently developed in~\cite{Venumadhav:2019tad,Venumadhav:2019lyq,Zackay:2019kkv}. We produce a comprehensive catalog of candidate events from our matched-filter search which covers binary neutron star (BNS), neutron star--black hole (NSBH), and binary black hole (BBH) mergers\footnote{\release}. While not individually significant on their own, sub-threshold candidates can be correlated with gamma-ray burst candidates~\citep{Nitz:2019bxt}, high-energy neutrinos~\citep{Countryman:2019pqq}, optical transients~\citep{Andreoni:2018fcm,Setzer:2018ppg}, and other counterparts to uncover new, fainter sources.

In addition to our broad search, we conduct a targeted analysis to uncover fainter BBH mergers. It is possible to confidently detect binary black hole mergers which are not individually significant in the context of the wider search space by considering their consistency with the population of confidently observed binary black hole mergers. The collection of highly significant detected events constrains astrophysical rates and distribution with relatively small uncertainties~\citep{LIGOScientific:2018jsj}. For this reason, we do not yet employ this technique for binary neutron star or neutron star--black hole populations, as their rates and mass and spin distributions are much less constrained.  We improve over the BBH focused analysis introduced in~\cite{Nitz:2018imz} by considering an explicit population prior~\citep{Dent:2013cva}. This focused approach is most directly comparable to the results of~\citep{Venumadhav:2019lyq}, which considers only binary black hole mergers, rather than a broad parameter search such as employed in~\cite{LIGOScientific:2018mvr}.

We find 8 highly significant binary black hole mergers at false alarm rates less than 1 per 100 years in our full analysis along with the binary neutron star merger, GW170817. No other individually significant BNS or NSBH sources were identified. However, if the population of these sources were to be better understood, it may be possible to pick out fainter mergers from our population of candidates. When we apply a ranking to search candidates that optimizes search sensitivity for a population of BBH mergers similar to that already detected, we identify a further 6 such mergers with a probability of astrophysical origin above $50\%$. These include GW170818 and GW170729 which were reported in~\cite{LIGOScientific:2018mvr} along with GW170121, GW170727, and GW170304 which were reported in~\cite{Venumadhav:2019lyq}. We report one new marginal BBH candidate, GW151205. Our results are broadly consistent with both~\cite{Venumadhav:2019lyq} and~\cite{LIGOScientific:2018mvr}.

\begin{table}[b]
  \begin{center}
    \caption{Observing time in days for different instrument observing combinations. We use here the abbreviations H, L, and V for the LIGO-Hanford, LIGO-Livingston, and
    Virgo observatories respectively. Note that some data ($\sim\!0.5\%$) may not be analyzed due to analysis constraints. Only the indicated combination of observatories were operating for each time period, hence each is exclusive of all others.}
    \label{table:data}
\begin{tabular}{lrllllll}
Observing & Time(days) \\\hline
H & 65.4 & \\
L & 50.0 & \\
V & 1.7 & \\ \hline
HL & 151.8 & \\
LV & 2.2 & \\
HV & 1.7 & \\ \hline
HLV & 15.2 & \\
\end{tabular}
  \end{center}
\end{table}

\section{LIGO and Virgo Observing Period}
We analyze the complete set of public LIGO and Virgo data~\citep{Vallisneri:2014vxa}. The distribution of multi-detector analysis time and the evolution of the observatories' sensitivities over time is shown in Table~\ref{table:data} and Fig.~\ref{fig:range} respectively. To date, there has been 288 days of Advanced LIGO and Virgo observing time. Two or more instruments were observing during 171 days. There were only 15.2 days of full LIGO-Hanford, LIGO-Livingston, and Virgo joint observing. O2 was the first time that Virgo has conducted joint observing with the LIGO interferometers since initial LIGO~\citep{Aasi:2013wya}. The Virgo instrument significantly surpassed the average BNS range during the last VSR2/3 science run ($\sim10$~Mpc)~\citep{Colaboration:2011np} to achieve an average of ~27~Mpc during the joint observing period of O2. While the amount of triple-detector observing time is limited during the first two observing runs, the ongoing third observing run will considerably improve the availability of three-detector joint observing time. The methods demonstrated here will be applicable to future analysis of the O3 multi-detector dataset. Our analysis during triple-detector time remains sensitive to signals which appear only in two of the three detectors, as discussed below in section~\ref{sec:multirank}.

We note that there is $\sim117$~days of single-detector observing time. In this work we do not consider the detection of gravitational-wave mergers during this time, however, methods for assigning meaningful significance to such events has been proposed~\citep{Callister:2017urp} and will be investigated in future work. Single-detector observing time has been used in follow-up analyses where a merger could be confirmed by electromagnetic observations~\citep{Nitz:2019bxt,Authors:2019fue}.

\section{Search for Binary Mergers}

We use a matched-filtering approach as implemented in the open source PyCBC library~\citep{Usman:2015kfa,Allen:2004gu,pycbc-github}. This toolkit has been similarly employed in LIGO/Virgo collaboration and independent analyses~\citep{LIGOScientific:2018mvr,Nitz:2018imz}. We extend the approach used in the 1-OGC analysis~\citep{Nitz:2018imz} to handle the analysis of Virgo data. We also incorporate improvements to the ranking of candidates by accounting for time variations in the power spectral density and network sensitivity.

The search procedure can be summarized as follows. The data from each detector is correlated against a set of possible merger signals. Matched filtering is used to calculate a signal-to-noise (SNR) time series for each potential signal waveform. Our analysis identifies peaks in these time series and follows up the peaks with a set of signal consistency tests. These single-detector candidates are then combined into multi-detector candidates by enforcing astrophysically consistent time delays between detectors, as well as enforcing identical component masses and spins. Finally, these candidates are ranked by the ratio of their signal and noise model likelihoods (see Sec.~\ref{sec:multirank}).
\begin{figure}[t]
  \centering
    \includegraphics[width=\columnwidth]{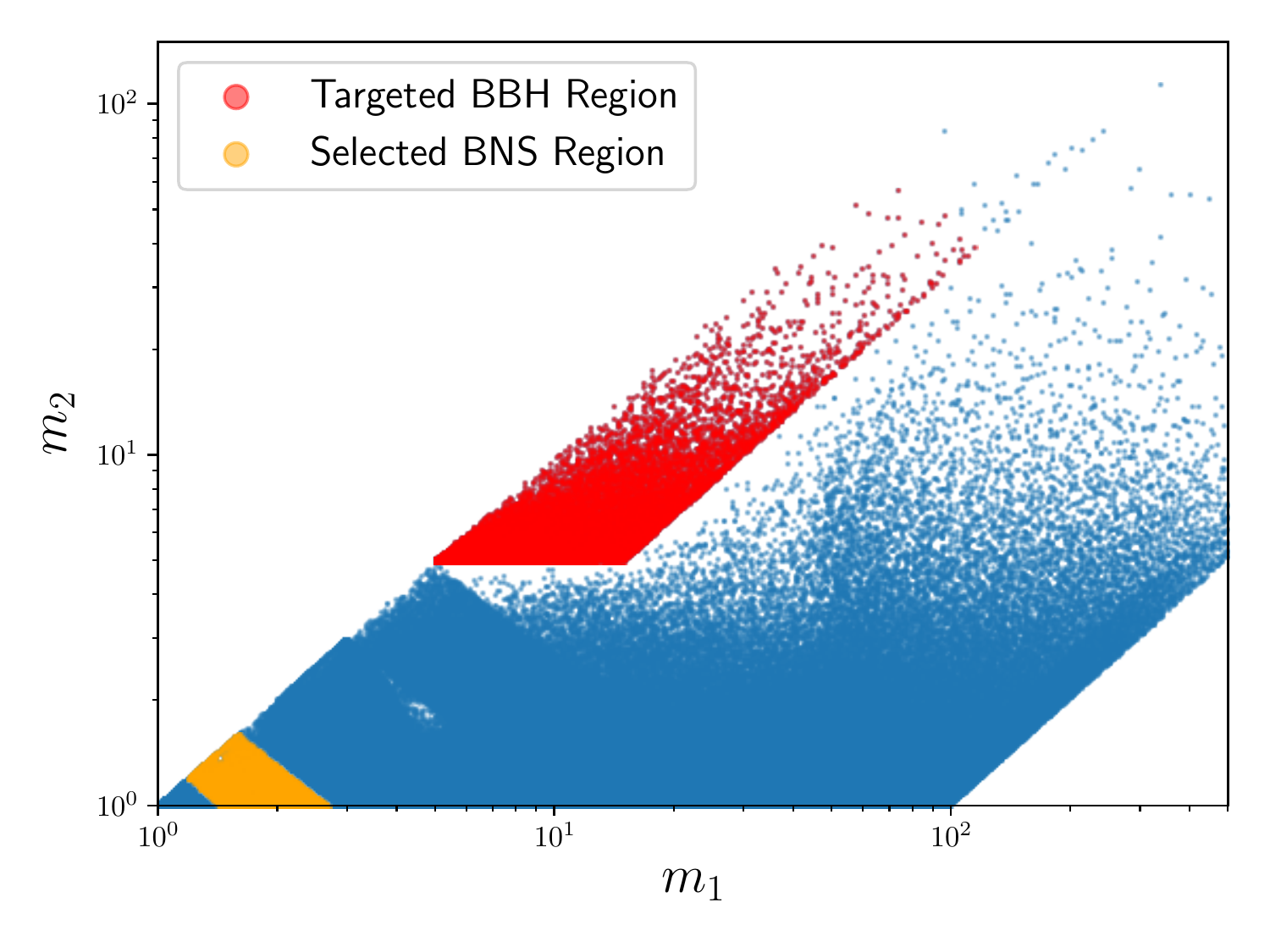}
\caption{The component masses of templates used to search for compact binary mergers. Templates in the targeted-binary black hole region, defined by $m_{1,2} > 5\,\msun$, $1/3 < m_1/m_2 < 3$, and $\mathcal{M}<60\,\msun$, where $\mathcal{M}$ is the chirp mass, are colored in red (see Sec.~\ref{sec:bbhsearch})}. Candidates which fall in the selected BNS-like region are discussed in~\ref{sec:nsobs}.
\label{fig:bank}
\end{figure}

\subsection{Search Space}
Our analysis targets a wide range of BNS, NSBH, and BBH mergers. We perform a matched filter on the data with waveform models that span the range of desired detectable sources. Although the space of possible binary component masses and spins is continuous, we must select a discrete set of points in this space as templates to correlate against the data: we use the set of $\sim 400,000$ templates introduced in~\cite{DalCanton:2017ala} which has been previously used in~\cite{Nitz:2018imz} and~\cite{LIGOScientific:2018mvr}. This bank of templates is suitable for the detection of mergers up to binary masses of several hundred solar masses, under the conditions that the dominant gravitational-wave emission mode is adequate to describe the signal and that the effects of precession caused by misalignment of the orbital and component object angular momenta can be neglected~\cite{DalCanton:2017ala}. Neglecting precession causes a 7\% (14\%) loss in sensitivity to BBH (NSBH) sources with mass ratio 5 (14) when assuming an isotropic distribution of the components' spins; the loss is negligible for mergers with comparable component masses~\citep{Harry:2016ijz}. Fig.~\ref{fig:bank} shows the distribution of template detector-frame component masses. We use the spinning effective-one-body model (SEOBNRv4) for templates corresponding to mergers with (redshifted, detector frame) total mass $m_1 + m_2 > 4\,M_\odot$~\citep{Taracchini:2013,Bohe:2016gbl}. The TaylorF2 post-Newtonian model is used in all other cases~\citep{Sathyaprakash:1991mt,Droz:1999qx,Blanchet:2002av,Faye:2012we}.

\subsection{Single-detector Candidates}

The first stage of our analysis is to identify single-detector candidates. These correspond to peaks in the SNR time series of a particular template waveform. Each is assigned a ranking statistic as we'll discuss below. In this work, we do not explicitly conduct a search for sources which only appear in a single detector. However, a ranking of single-detector candidates forms the first stage of our analysis. For each template waveform and detector dataset we calculate a signal-to-noise time series $\rho(t)$ using matched filtering. This can be expressed using a frequency-domain convolution as
\begin{equation}
    \rho(t) = 4 \Re\int_{f_l}^{f_h}{\frac{\tilde{h}^{*}(f)\tilde{s}(f)}{S_n(f)} e^{2\pi i f t} \, df},
\end{equation}
where $\tilde{h}$ is the normalized (Fourier domain) template waveform and $\tilde{s}$ is the detector data. $S_n$ is the noise power spectral density (PSD) of the data which is estimated
using Welch's method. The integration range extends from a template-dependent lower frequency limit $f_l$ (ranging from 20-30Hz in our search) to an upper cutoff $f_h$ given by the Nyquist frequency of the data.
Peaks in the SNR time series are collected as single-detector candidates (triggers). To control the rate of single-detector candidates to be examined, our analysis pre-clusters these triggers. Only those which are amongst the loudest 100 every $\sim1$\,s within a set of predefined chirp-mass bins are kept. The binning ensures that loud triggers from a specific region (which may be caused by non-Gaussian noise artifacts) do not cause quiet signals elsewhere in parameter space to be missed.

We remove candidates where the instrument state indicates the data may be adversely affected by instrumental or environment noise artefacts as indicated by the Gravitational-Wave Open Science Center (GWOSC)~\citep{TheLIGOScientific:2016zmo,TheLIGOScientific:2017lwt,Vallisneri:2014vxa}. This affects $\sim0.5\%$ of the observation period. However, there remain classes of transient non-Gaussian noise in the LIGO data which produce triggers with large values of SNR~\citep{Nuttall:2015dqa,TheLIGOScientific:2016zmo,TheLIGOScientific:2017lwt,Cabero:2019orq}.
The surviving single-detector candidates are subjected to the signal consistency tests introduced in~\cite{Allen:2004gu} and~\cite{Nitz:2017lco}. These tests
check that the accumulation of signal power as a function of frequency, and power outside the expected signal band, respectively, are consistent with an astrophysical explanation. They produce two statistic values which are $\chi^2$ distributed: $\chi_r^2$ and $\chi_{r,sg}^2$ respectively~\citep{Nitz:2017lco}. These are used to re-weight~\citep{Babak:2012zx} the single-detector signal strength in two stages. This re-weighting allows candidates which well match an expected astrophysical source to be assigned a statistic value similar to their matched filter SNR, while down-weighting many classes of non-Gaussian noise transients. For all candidates we apply
\begin{equation}
 \tilde{\rho} = \begin{cases}
        \rho, & \mathrm{for}\ \chi^2_r \leq 1 \\
        \rho\left[ \frac{1}{2} \left(1 + \left(\chi^2_r\right)^3\right)\right]^{-1/6}, &
        \mathrm{for}\ \chi^2_r > 1.
    \end{cases}
\end{equation}
For single-detector candidates in templates with (detector-frame) total mass greater than $40\,\msun$ the statistic is further re-weighted as
\begin{equation}
\label{eq:sg}
 \hat{\rho} = \begin{cases}
        \tilde{\rho}, & \mathrm{for}\ \chi^2_{r,sg} \leq 4 \\
        \tilde{\rho} (\chi^2_{r,sg} / 4)^{-1/2}, &
        \mathrm{for}\ \chi^2_{r,sg} > 4.
    \end{cases}
\end{equation}
The latter test is only applied to these short duration, higher mass signals as the test is computationally intensive and has the greatest impact for short duration signals which may be otherwise confused with some classes of transient non-Gaussian noise~\citep{Nitz:2017lco,Cabero:2019orq}.  Otherwise, we set $\hat{\rho} = \tilde{\rho}$.

This statistic $\hat{\rho}$ is the same used in the 1-OGC analysis~\citep{Nitz:2018imz} and LVC O2 catalog \cite{LIGOScientific:2018mvr}. We further improve upon this by accounting for short-term changes in the overall PSD estimate. The issue of PSD variation was also addressed in \cite{Venumadhav:2019tad}.
Previously we modeled the PSD for each detector as a function of frequency $S_n(f)$, which is estimated on a 512 second timescale. We now introduce a time-dependent factor $v_S(t)$ which accounts for short-term O(10s) variations in sensitivity, estimated using the method described in \cite{Simone:2019prep}. Short-term variation in the PSD will introduce variation in $\rho$ as we use the estimated PSD $S(f)$ to calculate it. The estimated PSD over short time scales $S_s(f)$ can be different from a PSD estimated over a longer duration $S_l(f)$ if the noise is non-stationary.

To track the variation in the PSD we use the variance of the SNR. In the absence of a signal, this is given by
\begin{equation}
    \langle\rho^2\rangle = \left[\int_{f_l}^{f_h}\frac{|\tilde{h}(f)|^2}{S_l(f)}\, df\right]^{-1} \int_{f_l}^{f_h}\frac{|\tilde{h}(f)|^2}{S_l(f)}\frac{S_s(f)}{S_l(f)}\, df.
\end{equation}
The variance is equal to $1$ if $S_s(f) = S_l(f)$.
To estimate the variance of the SNR, we first filter the detector data $\tilde{s}$ with $F(f) = \mathcal{N}|\tilde{h}(f)|/S_l(f)$, where $\mathcal{N}$ is a normalization constant and $|\tilde{h}(f)|$ is an approximation to the Fourier domain amplitude of CBC templates. Using Parseval's theorem, we can then estimate the variation in the PSD at a given time $t_0$ as
\begin{equation}
    v_S(t_0) = \int_{t_0-\Delta t}^{t_0}|\mathcal{F}(t)*s(t)|^2\, dt,
\end{equation}
where $\mathcal{F}(t)*s(t)$ is the convolution between the filter and the data and $\Delta t$ is chosen to match the typical time scale of non-stationarity.

After finding $v_S(t)$, we evaluate its correlation with the SNRs and rates of noise triggers empirically.  The rate of noise triggers above a given statistic threshold is $R_N(>\hat{\rho}_t)$, where the statistic $\hat{\rho}$ is (proportional to) the SNR obtained by matched filtering using the long-duration PSD $S_l(f)$.  The noise trigger rate varies over time due to the non-stationarity of the PSD and is thus a function of the short-duration PSD variation measure.  Since SNR scales as $1/\sqrt{S(f)}$, we naively expect the noise trigger rate to be a function of a `corrected' SNR $\hat{\rho}/\sqrt{v_S}$; in practice we allow for a more general dependence which we write as
\begin{equation}
    R_N(>\hat{\rho}_t) \simeq f_N(\hat{\rho}_t v_S(t)^{-\kappa}).
\end{equation}
Here $f_N$ is a fitting function for the expected noise distribution.  Empirically we find, for data without strong localized non-Gaussian transients (glitches), $f_N(\hat{\rho}) \simeq \exp(-\alpha \hat{\rho}^2/2)$ with $\alpha \simeq 1$.

Linearizing the PSD variation measure $v_S(t)$ around unity, $v_S(t) = 1 + \epsilon_S$, the logarithm of the trigger rate above threshold will vary as
\begin{equation}
    \ln(R_N(>\hat{\rho}_t)) \simeq \mathrm{const.} -
    \frac{\alpha}{2}\hat{\rho}_t^2 +
    \alpha\kappa \hat{\rho}_t^2 \epsilon_S
\end{equation}
By determining the slope of the log-rate versus\ $\epsilon_S$ dependence for various thresholds $\hat{\rho}$ we estimate $\kappa\sim 0.33$, thus if we construct a `corrected' statistic
\begin{equation}
     \hat{\rho} \xrightarrow{} \hat{\rho} v_S(t)^{-0.33},
\end{equation}
the rate of noise triggers above a given threshold of the corrected statistic is on average no longer affected by variation in $v_S(t)$.

The analysis of \cite{Venumadhav:2019tad} included a similar correction factor and \cite{Zackay:2019kkv} indicates a modest improvement in sensitivity for the sources they consider. In our analysis, the greatest improvement is for sources corresponding to long-duration templates (BNS and NSBH) while there is negligible improvement for the shorter-duration BBH sources.

\subsection{Multi-detector Coincident Candidates}
\label{sec:multirank}

In the previous section, we discussed how we identify single-detector candidates and assign them a ranking statistic. We now combine single-detector candidates from multiple detectors to form multi-detector candidates~\citep{Davies:2020}. We introduce a new ranking statistic formed from models of the relative signal and noise likelihoods for a particular candidate. This ranking statistic is based on the expected rates of signal and noise candidates, and is thus comparable across different combinations of detectors by design. We are then able to search for coincident triggers in all available combinations of detectors (for instance during HLV time, coincidences can be formed in HL, HV, LV and HLV), and then compared to one another, clustering and combining false alarm rates while maintaining near-optimal sensitivity.

Our signal model is composed of two parts. First, the overall network sensitivity of the analysis at the time of the candidate. Assuming a spatially homogeneous distribution of sources, the signal rate is directly proportional to the sensitive volume. We approximate this factor using the instantaneous range of the least sensitive instrument contributing to the multi-detector candidate for a given template labelled by $i$, $\sigma_{{\rm min}, i}$, relative to a representative range over the analysis, which is defined by the median network sensitivity in the HL detector network, $\overline\sigma_{HL, i}$, for that template. Note that the detectors which contribute to the candidate are not necessarily all of the available detectors at that time.
The second part is the probability, given an isotropic and volumetric population of sources, that an astrophysical signal would be observed to have a particular set of parameters defined by $\vec{\theta}$, including time delays, relative amplitudes, and relative phases between the network of observatories.
This probability distribution $p(\vec{\theta}|S)$ is calculated
by a Monte Carlo method similarly to~\cite{Nitz:2017svb}. For this work we have extended this technique to three detectors for the first time. Combined, our model for the density of signals recovered with network parameters $\vec{\theta}$ in a combination of instruments characterized by $\sigma_{min, i}$ can be expressed as
\begin{equation}
    R_{S,i}(\vec{\theta}) = \left(\frac{\sigma_{{\rm min}, i}}{\overline\sigma_{HL, i}}\right)^3 p(\vec{\theta}|S) .
\end{equation}

The noise model is calculated in the same manner as in ~\cite{Nitz:2017svb}. We treat the noise from each detector as being independent and fit our single-detector ranking statistic to an exponential slope. This fit is performed separately for each template. The fit parameters (such as the slope and overall amplitude of the exponential) are initially noisy due to low number statistics, so they are smoothed over the template space using a three-dimensional Gaussian kernel in the template duration, effective spin $\chieff$, and symmetric mass ratio $\eta$ parameters. The rate density of noise events in the $i^\textrm{th}$ template with contributing detectors labelled by $n$ and single-detector rankings $\{\tilde{\rho}_{n}\}$ can be summarized as
\begin{equation}
    R_{N,i}(\{\tilde{\rho}_{n}\}) = A_{\{n\}}\prod_{n}{r_{n, i}e^{-\alpha_{n,i} \tilde{\rho}_{n}}},
\end{equation}
where $r_{n,i}$ and $\alpha_i$ are the overall amplitude and slope of the exponential noise rate model respectively.
The \mbox{prefactor} $A_{\{n\}}$ is the time window for which coincidences can be formed, which depends on the combination of detectors $\{n\}$ being considered. The three-detector coincidence rate is vastly reduced compared to the two-detector rate; in a representative stretch of O2 data, the HLV coincidence rate is found to be around a factor of $10^4$ lower than that in HL coincidences. Details of both the signal and noise model calculations will be provided in~\cite{Davies:2020}.

The ranking statistic for a given candidate in template $i$ is the log of the ratio of these two rate densities:
\begin{equation}
    \rankingstat = \log(R_{S,i} / R_{N,i}) + \mathrm{const.},
\end{equation}
where we drop the dependencies on $\vec{\theta}$ and $\{\tilde{\rho}_{n}\}$ for simplicity of notation.  Typically, one signal event (or loud noise event) in the gravitational-wave data stream may give rise to a large number of correlated candidate multi-detector events within a short time, in different templates and with different combinations of detectors $\{n\}$.  To calculate the significance of such a `cluster' of events, we will approximate their arrival as a Poisson process: in order to do this, we keep the event from each cluster with highest $\rankingstat$ --- typically the highest-ranked event within a 10\,s time window --- and discard the rest.

Comparing this new statistic to the one employed for the 1-OGC analysis~\citep{Nitz:2018imz} using a simulated population of mergers, we find an average $8\%$ increase in the detectable volume during the O1 period at a fixed false alarm rate of 1 per 100 years. This population is isotropically distributed in sky location and orientation, while the mass distribution is scaled to ensure a constant rate of signals above a fixed SNR across the log-component-mass search space in Fig.~\ref{fig:bank}. In addition to this improved sensitivity for events where H1 and L1 contribute, this search will also benefit by analyzing times where Virgo and only one LIGO detector are operating (as in Table~\ref{table:data}), and also by improved sensitivity in times when all three detectors are operating, due to the ability to form 3-detector events.  Such sensitivity improvements are detailed in~\cite{Davies:2020}.

\subsection{Statistical Significance}

In the previous section we introduce the ranking statistic used in our analysis. We empirically measure the statistical significance of a particular value of our ranking statistic by comparing it to a set of false (noise) candidate events produced in numerous fictitious analyses. Each such analysis is generated by time-shifting the data from one detector by an amount greater than is astrophysically allowed by light travel time considerations~\cite{Babak:2012zx,Usman:2015kfa}. Otherwise, each time-shifted analysis is treated in the identical manner as the search itself. By repeating this procedure, upwards of $10^4$ years' worth of false alarms can be produced from just a few days of data. By construction, the results of these analyses cannot contain true multi-detector astrophysical candidates, but may contain coincidences between astrophysical sources and instrumental noise. We use a hierarchical procedure as in \cite{TheLIGOScientific:2016pea} and \cite{Nitz:2018imz} to minimize the impact of astrophysical contamination while retaining an unbiased rate of false alarms~\citep{2017PhRvD..96h2002C}: a candidate with large \rankingstat{} is removed from the estimation of background for less significant candidates.

This method has been employed to detect significant events in numerous analyses~\citep{Nitz:2018imz,LIGOScientific:2018mvr,Venumadhav:2019lyq,Colaboration:2011np,Abbott:2009qj}. The validity of the resulting background estimate follows from an assumption that the times of occurrence of noise events are statistically independent between different detectors; see \cite{Was:2009vh,2017PhRvD..96h2002C} for further discussion of empirical background estimation and the time shift method. This is a reasonable assumption for detectors separated by thousands of kilometres~\citep{TheLIGOScientific:2016zmo}. The time shift method has the advantage that no other assumptions about the noise need be accurate: the populations and morphology of noise artefacts need not be uncorrelated or different between detectors, only the times at which they occur.  In fact the LIGO and Virgo instruments share common components and environmental coupling mechanisms which may produce similar classes of non-Gaussian artefacts.
\begin{table*}
  \begin{center}
    \caption{Candidate events in the full search of O1 and O2 data. Candidates are sorted by FAR evaluated for the entire bank of templates. Note that ranking statistic and false alarm rate may not have a strictly monotonic relationship due to varying data quality between sub-analyses.  The mass and spin parameters listed are associated with the template waveform yielding the highest ranked multi-detector event for each candidate, and may differ significantly from full Bayesian parameter estimates. Masses are quoted in detector frame, and are thus larger than source frame masses by a factor $(1+z)$, where $z$ is the source redshift.}
    \label{table:complete}
\begin{tabular}{lcrcrrrrrrrr}
Date designation & GPS time & FAR$^{-1}$ (y) & Detectors & $\rankingstat$ & $\rho_H$ & $\rho_L$ & $\rho_V$ & $m_1^t$ & $m_2^t$ & $\chieff^t$\\ \hline
170817+12:41:04UTC\gwtc         & 1187008882.45          &   $>10000$          &         HL          &     180.46          &       18.6          &       24.3          &          -          &        1.5          &        1.3          &    -0.00\\
150914+09:50:45UTC\gwtc\ogc\ias          & 1126259462.43          &   $>10000$          &         HL          &      93.82          &       19.7          &       13.4          &          -          &       44.2          &       32.2          &     0.09\\
170104+10:11:58UTC\gwtc\ias          & 1167559936.60          &   $>10000$          &         HL          &      35.54          &        9.0          &        9.6          &          -          &       47.9          &       16.0          &     0.03\\
170823+13:13:58UTC\gwtc\ias          & 1187529256.52          &   $>10000$          &         HL          &      55.04          &        6.3          &        9.2          &          -          &       68.9          &       47.2          &     0.23\\
170814+10:30:43UTC\gwtc\ias          & 1186741861.54          &   $>10000$          &         HL          &      52.85          &        9.0          &       13.0          &          -          &       58.7          &       23.3          &     0.53\\
151226+03:38:53UTC\gwtc\ogc\ias          & 1135136350.65          &   $>10000$          &         HL          &      42.90          &       10.7          &        7.4          &          -          &       14.8          &        8.5          &     0.24\\
170809+08:28:21UTC\gwtc\ias          & 1186302519.76          &       9400          &         HL          &      40.59          &        6.6          &       10.7          &          -          &       36.0          &       33.7          &     0.07\\
170608+02:01:16UTC\gwtc          & 1180922494.49          &        $>910$\footnote{\label{170608}The FAR is limited only by the available background data. A short analysis period is used for the 170608 data which was released separately due to an instrument angular control procedure affecting data from the Hanford observatory~\citep{Abbott:2017gyy}.}            &         HL          &      51.01          &       12.5          &        8.7          &          -          &       16.8          &        6.1          &     0.31\\
151012+09:54:43UTC\gwtc\ogc\ias          & 1128678900.45          &        220          &         HL          &      20.18          &        7.0          &        6.7          &          -          &       30.8          &       12.9          &    -0.05\\
170729+18:56:29UTC\gwtc\ias          & 1185389807.33          &        6.4          &         HL          &      15.33          &        7.4          &        6.7          &          -          &      106.5          &       49.7          &     0.59\\
170121+21:25:36UTC\ias          & 1169069154.58          &        1.3          &         HL          &      15.76          &        5.1          &        8.7          &          -          &       40.4          &       13.6          &    -0.98\\
170727+01:04:30UTC\ias          & 1185152688.03          &        .53          &         HL          &      13.75          &        4.5          &        6.9          &          -          &       65.2          &       26.5          &    -0.35\\
170818+02:25:09UTC\gwtc          & 1187058327.09          &        .22          &         HL          &      13.29          &        4.4          &        9.4          &          -          &       53.7          &       27.4          &     0.07\\
170722+08:45:14UTC          & 1184748332.91          &        .11          &         HL          &      12.19          &        5.0          &        6.4          &          -          &      248.1          &        7.1          &     0.99\\
170321+03:13:21UTC          & 1174101219.23          &         .1          &         HL          &      12.22          &        6.5          &        6.4          &          -          &       11.0          &        1.3          &    -0.89\\
170310+09:30:52UTC          & 1173173470.77          &        .07          &         HL          &      12.15          &        6.1          &        6.2          &          -          &        2.1          &        1.1          &    -0.20\\
170809+03:55:52UTC          & 1186286170.08          &        .07          &         LV          &       7.34          &          -          &        7.0          &        5.1          &        6.2          &        1.2          &     0.60\\
170819+07:30:53UTC          & 1187163071.23          &        .05          &         HV          &      11.35          &        6.3          &          -          &        6.7          &      135.2          &        2.5          &     0.85\\
170618+20:00:39UTC          & 1181851257.72          &        .05          &         HL          &      11.49          &        5.2          &        6.7          &          -          &        2.9          &        2.1          &     0.30\\
170416+18:38:48UTC          & 1176403146.15          &        .04          &         HL          &      11.21          &        5.1          &        6.9          &          -          &        7.8          &        1.1          &    -0.47\\
170331+07:08:18UTC          & 1174979316.31          &        .04          &         HL          &      11.03          &        5.2          &        7.0          &          -          &        3.9          &        1.1          &    -0.34\\
151216+18:49:30UTC          & 1134326987.60          &        .04          &         HL          &      11.54          &        6.1          &        6.0          &          -          &       13.9          &        5.0          &    -0.41\\
170306+04:45:50UTC          & 1172810768.08          &        .04          &         HL          &      11.47          &        4.8          &        7.3          &          -          &       26.4          &        1.8          &     0.23\\
151227+16:52:22UTC          & 1135270359.27          &        .04          &         HL          &      11.75          &        7.3          &        4.6          &          -          &      154.5          &        4.9          &     1.00\\
170126+23:56:22UTC          & 1169510200.17          &        .04          &         HL          &      11.61          &        6.4          &        5.7          &          -          &        4.9          &        1.3          &     0.79\\
151202+01:18:13UTC          & 1133054310.55          &        .03          &         HL          &      11.48          &        6.5          &        5.7          &          -          &       40.4          &        1.8          &    -0.26\\
170208+20:23:00UTC          & 1170620598.15          &        .03          &         HL          &      11.12          &        6.8          &        5.4          &          -          &        6.9          &        1.0          &     0.09\\
170327+17:07:35UTC          & 1174669673.72          &        .03          &         HL          &      10.65          &        6.0          &        6.2          &          -          &       40.1          &        1.0          &     0.97\\
170823+13:40:55UTC          & 1187530873.86          &        .03          &         LV          &       9.30          &          -          &        8.0          &        5.8          &      117.9          &        1.3          &     0.98\\
150928+10:49:00UTC          & 1127472557.93          &        .03          &         HL          &      11.28          &        6.0          &        6.3          &          -          &        2.5          &        1.0          &    -0.70\\
\end{tabular}
  \end{center}
\footnotesize{\label{prev}$^x$Also identified in GWTC-1~\citep{LIGOScientific:2018mvr}, $^y$1-OGC~\citep{Nitz:2018imz}, or $^z$\cite{Venumadhav:2019tad,Venumadhav:2019lyq}}
\end{table*}

\subsection{Targeting Binary Black Hole Mergers}
\label{sec:bbhsearch}
Given a population of individually significant BBH mergers, it is possible to incorporate knowledge about the overall distribution and rate of sources to identify weaker candidates. A similar approach was employed in~\cite{Nitz:2018imz} and is the basis of astrophysical significance statements in~\cite{LIGOScientific:2018mvr}. In this catalog we improve over the strategy of~\cite{Nitz:2018imz} which considered an excessively conservative parameter space for BBH and did not use an explicit model of the distribution of signals and noise within that space. In addition, we restrict to sources which are consistent with our signal models by imposing a threshold on our primary signal consistency test to reject any single-detector candidate with $\chi_r > 2.0$. Simulated signals within our target population, and the individual highly significant candidates previously detected are consistent with this choice.  (The full, non-BBH-specific analysis allows a much greater deviation from our signal models before rejection of a candidate.)

As a first step in obtaining the targeted BBH results we restrict the analysis to a sub-space of the full search, illustrated in Fig.~\ref{fig:bank}.  Rather than applying this constraint \emph{after} obtaining the set of `clustered' candidates via selecting the highest ranked event within $10$\,s windows, as in~\cite{Nitz:2018imz}, here we apply the constraint to candidates \emph{prior} to the clustering step.  This allows us to choose a less extensive BBH region containing fewer templates than employed in~\cite{Nitz:2018imz} without loss of sensitivity.  (The previous method used a wider BBH template set to allow for the possibility that a signal inside the intended target region is recovered only by a template lying outside that region, due to clustering.)
Our BBH region is specified by $m_{1,2} > 5\,\msun$, $1/3 < m_1/m_2 < 3$, and $\mathcal{M}<60\,\msun$. The upper boundary is consistent with the redshifted detector-frame masses that would be obtained by the observed highest-mass sources near detection threshold.

Applying a prior over the intrinsic parameters of the distribution of detectable sources was proposed in~\cite{Dent:2013cva} and tested in~\cite{Nitz:2017svb}. In this work, we impose an explicit detection prior that is flat over chirp mass. As seen in Fig.~\ref{fig:bank}, the distribution of templates is highly non-uniform. The BBH region of template is placed first using a stochastic algorithm~\citep{DalCanton:2017ala,Ajith:2012mn}, where density of templates directly correlates to density of effectively independent noise events. The template density over $\mathcal{M}$ scales as $\mathcal{M}^{-11/3}$, which we verify empirically for our bank.  Our detection statistic aims to follow the relative rate density of signal versus noise events at fixed SNR, and we make the simple choice of assuming a signal density flat over $\mathcal{M}$: thus the ranking statistic receives an extra term describing the ratio of signal to noise densities over component masses:
\begin{equation}
    \rankingstat{}_{\rm BBH} = \rankingstat{} + \frac{11}{3}\ln\left(\frac{\mathcal{M}}{\mathcal{M}_f}\right),
\end{equation}
where $\mathcal{M}_f = 20\,\msun$ is a fiducial reference mass scale.
Roughly, any given lower-mass template is less likely to detect a signal than a higher-mass template given that templates are much sparser at high masses.

Our choice of BBH region and detection prior has a similar effect as the highly constrained search space and multiple chirp mass bins used in~\cite{Venumadhav:2019lyq} but avoids the multiple boundary effects present there and provides a more clearly implemented and astrophysically motivated prior distribution.  Furthermore, our method provides a path forward to more accurate assessment of lower-SNR candidates as our understanding of the overall population evolves.

To estimate the probability \pastro\ that a given candidate is astrophysical in origin we combine the background of this targeted BBH analysis with the estimated distribution of observations. We improve upon the analysis in~\cite{Nitz:2018imz} which employed an analytic model of the signal distribution and a fixed conservative rate of mergers by using the mixture model method developed in~\cite{Farr:2015} and similar to that employed in~\cite{LIGOScientific:2018mvr}.
This method requires the distribution of noise and signals over our ranking statistic, which we take from our time-slide background estimates and a population of simulated signals respectively.\footnote{We use the Laguerre-Gauss integral method described in \cite{CreightonIdentities} to marginalize over the Poisson rate of signals in the calculation of \pastro\ values.}

Using a simulated set of mergers that is isotropically distributed in orientation and uniformly distributed over mass to cover the targeted BBH region, we find that the targeted BBH analysis recovers a factor $1.5$--$1.6$ more sources at a fixed false alarm rate of 1 per 100 years than the full parameter space analysis. The majority of this change in sensitivity is attributed to the inclusion of only background events consistent with BBH mergers. The choice of ranking statistic to optimize sensitivity to a target BBH signal population has a smaller effect.

\renewcommand{\ias}{\hyperref[prev2]{\textsuperscript{z}}}
\renewcommand{\gwtc}{\hyperref[prev2]{\textsuperscript{x}}}
\renewcommand{\ogc}{\hyperref[prev2]{\textsuperscript{y}}}

\begin{table*}
  \begin{center}
    \caption{Candidates from the targeted binary black hole sub-region sorted by the probability they are astrophysical in origin. The source-frame masses, \chieff, and luminosity distance $D_L$ are estimated with Bayesian parameter inference (see Sec.~\ref{sec:bbh}) and are given with $90\%$ credible intervals.}
    \label{table:bbh}
\hspace*{-4cm}
\resizebox{2.7\columnwidth}{!}{%
\begin{tabular}{lcrrcrrrrllll}
Date designation & GPS time & \pastro & FAR$^{-1}$ (y) & Det. & $\rankingstat_{\rm BBH}$ & $\rho_H$ & $\rho_L$ & $\rho_V$ & $m^{\mathrm{src}}_{1}$ & $m^{\mathrm{src}}_{2}$ & $\chieff$ & $D_L$ (Mpc) \\ \hline
150914+09:50:45UTC\gwtc\ogc\ias          &  1126259462.43           &    $>0.999$           &    $>10000$           &          HL           &      111.71           &        19.7           &        13.4           &           -           & $35.4^{+5.3}_{-3.2}$ & $29.8^{+3.1}_{-4.7}$ & $-0.04^{+0.11}_{-0.13}$ & $470^{+140}_{-190}$\\
170814+10:30:43UTC\gwtc\ias           &  1186741861.53           &    $>0.999$           &    $>10000$           &          HL           &       61.58           &         9.3           &        13.8           &           -           & $30.4^{+5.6}_{-2.7}$ & $25.8^{+2.6}_{-4}$ & $0.08^{+0.12}_{-0.12}$ & $580^{+130}_{-190}$\\
170823+13:13:58UTC\gwtc\ias           &  1187529256.52           &    $>0.999$           &    $>10000$           &          HL           &       59.43           &         6.3           &         9.2           &           -           & $40^{+11.7}_{-7.1}$ & $28.8^{+6.8}_{-7.9}$ & $0.05^{+0.21}_{-0.22}$ & $1750^{+850}_{-820}$\\
170104+10:11:58UTC\gwtc\ias           &  1167559936.60           &    $>0.999$           &    $>10000$           &          HL           &       47.32           &         9.1           &         9.9           &           -           & $31.6^{+7.8}_{-6.3}$ & $19.2^{+5}_{-4.1}$ & $-0.08^{+0.16}_{-0.18}$ & $920^{+420}_{-400}$\\
151226+03:38:53UTC\gwtc\ogc\ias           &  1135136350.65           &    $>0.999$           &    $>10000$           &          HL           &       40.58           &        10.7           &         7.4           &           -           & $13.9^{+7.9}_{-3.3}$ & $7.6^{+2.2}_{-2.3}$ & $0.209^{+0.177}_{-0.077}$ & $460^{+160}_{-180}$\\
151012+09:54:43UTC\gwtc\ogc\ias           &  1128678900.45           &    $>0.999$           &    $>10000$           &          HL           &       20.25           &         7.0           &         6.7           &           -           & $22.4^{+13.4}_{-4.8}$ & $13.8^{+3.7}_{-4.8}$ & $-0.00^{+0.25}_{-0.16}$ & $990^{+470}_{-460}$\\
170809+08:28:21UTC\gwtc\ias           &  1186302519.76           &    $>0.999$           &        8300\footnote{\label{farlimit}The false alarm rate is limited by false coincidences arising from the candidate's time-shifted LIGO-Livingston single-detector trigger. If removed from its own background, the FAR is $<$ 1 per 10,000 years.}             &          HL           &       43.34           &         6.6           &        10.7           &           -           & $35.2^{+9.5}_{-5.9}$ & $23.9^{+5.1}_{-5.3}$ & $0.06^{+0.18}_{-0.16}$ & $980^{+310}_{-390}$\\
170729+18:56:29UTC\gwtc\ias           &  1185389807.33           &    $>0.999$           &        4000           &          HL           &       19.16           &         7.5           &         7.1           &           -           & $55^{+18}_{-13}$ & $32^{+13}_{-10}$ & $0.31^{+0.22}_{-0.29}$ & $2300^{+1600}_{-1300}$\\
170608+02:01:16UTC\gwtc           &  1180922494.49           &    $>0.999$           &         $>910$        &          HL           &       55.12           &        12.5           &         8.7           &           -           & $11.6^{+6.7}_{-2.1}$ & $7.4^{+1.6}_{-2.3}$ & $0.088^{+0.213}_{-0.073}$ & $310^{+130}_{-110}$\\
170121+21:25:36UTC\ias           &  1169069154.58           &    $>0.999$           &         210\textsuperscript{\ref{farlimit}}  &          HL           &       23.86           &         5.1           &         8.9           &           -           & $33^{+9.2}_{-5.3}$ & $25.7^{+5.3}_{-6.1}$ & $-0.17^{+0.24}_{-0.26}$ & $1150^{+950}_{-650}$\\
170818+02:25:09UTC\gwtc           &  1187058327.09           &    $>0.999$           &         5.1\textsuperscript{\ref{farlimit}}          &          HL           &       21.42           &         4.4           &         9.4           &           -           & $36^{+8.2}_{-5.3}$ & $26.2^{+4.8}_{-5.7}$ & $-0.11^{+0.20}_{-0.23}$ & $980^{+430}_{-340}$\\
170727+01:04:30UTC\ias           &  1185152688.03           &       0.994           &         180           &          HL           &       15.84           &         4.5           &         6.9           &           -           & $41.6^{+12.8}_{-7.9}$ & $30.4^{+7.9}_{-8.2}$ & $-0.05^{+0.25}_{-0.30}$ & $2200^{+1500}_{-1100}$\\
170304+16:37:53UTC\ias           &  1172680691.37           &        0.70           &         2.5           &          HL           &       11.61           &         4.6           &         7.1           &           -           & $44.9^{+17.6}_{-9.4}$ & $31.8^{+9.5}_{-11.6}$ & $0.11^{+0.29}_{-0.27}$ & $2300^{+1600}_{-1200}$\\
151205+19:55:25UTC           &  1133380542.42           &        0.53           &         .61           &          HL           &       10.97           &         5.8           &         4.8           &           -           & $67^{+28}_{-17}$ & $42^{+16}_{-19}$ & $0.14^{+0.40}_{-0.38}$ & $3000^{+2400}_{-1600}$\\
151217+03:47:49UTC           &  1134359286.35           &        0.26           &         .15           &          HL           &        9.61           &         6.7           &         5.6           &           -           & $46^{+13}_{-26}$ & $8.2^{+5.1}_{-1.7}$ & $0.70^{+0.15}_{-0.50}$ & $1000^{+660}_{-440}$\\
170201+11:03:12UTC           &  1169982210.74           &        0.24           &         .16           &          HL           &        9.26           &         6.0           &         5.6           &           -           & $48^{+13}_{-23}$ & $13.1^{+8.6}_{-3.7}$ & $0.44^{+0.28}_{-0.54}$ & $1530^{+1360}_{-770}$\\
170425+05:53:34UTC\ias           &  1177134832.19           &        0.21           &          .2           &          HL           &        9.42           &         5.1           &         5.8           &           -           & $45^{+21}_{-11}$ & $30^{+11}_{-11}$ & $-0.06^{+0.28}_{-0.32}$ & $2600^{+2000}_{-1300}$\\
151216+09:24:16UTC\ogc\ias           &  1134293073.19           &        0.18           &          .1           &          HL           &        9.25           &         5.9           &         5.5           &           -           & $41^{+15}_{-17}$ & $14.4^{+7}_{-6.3}$ & $0.51^{+0.21}_{-0.57}$ & $1620^{+1140}_{-910}$\\
170202+13:56:57UTC\ias           &  1170079035.73           &        0.13           &         .06           &          HL           &        8.37           &         5.0           &         6.6           &           -           & $33^{+17}_{-11}$ & $13.8^{+7}_{-4.8}$ & $-0.06^{+0.27}_{-0.32}$ & $1220^{+980}_{-640}$\\
170104+21:58:40UTC           &  1167602338.72           &        0.12           &         .03           &          HL           &        8.80           &         5.6           &         5.4           &           -           & $98^{+49}_{-40}$ & $44^{+30}_{-33}$ & $0.25^{+0.50}_{-0.49}$ & $4600^{+4300}_{-3100}$\\
170220+11:36:24UTC           &  1171625802.53           &        0.10           &         .05           &          HL           &        8.43           &         4.4           &         5.2           &           -           & $69^{+37}_{-25}$ & $31^{+22}_{-14}$ & $0.28^{+0.33}_{-0.37}$ & $3600^{+3700}_{-2100}$\\
170123+20:16:42UTC           &  1169237820.55           &        0.08           &         .04           &          HL           &        7.97           &         5.0           &         5.3           &           -           & $44^{+23}_{-12}$ & $28^{+13}_{-13}$ & $-0.12^{+0.31}_{-0.35}$ & $2800^{+2800}_{-1600}$\\
151011+19:27:49UTC           &  1128626886.61           &        0.08           &         .12           &          HL           &        8.45           &         4.9           &         6.6           &           -           & $51^{+18}_{-12}$ & $31^{+12}_{-12}$ & $0.09^{+0.29}_{-0.27}$ & $1560^{+1090}_{-740}$\\
151216+18:49:30UTC           &  1134326987.60           &        0.07           &         .03           &          HL           &        8.14           &         6.1           &         6.0           &           -           & $19.7^{+6.4}_{-7.4}$ & $3.25^{+1.32}_{-0.58}$ & $-0.03^{+0.24}_{-0.49}$ & $500^{+280}_{-250}$\\
170721+05:55:13UTC           &  1184651731.37           &        0.06           &         .04           &          HL           &        7.76           &         6.6           &         5.1           &           -           & $31.7^{+9.3}_{-6.1}$ & $21.4^{+5.3}_{-5.6}$ & $-0.06^{+0.25}_{-0.29}$ & $1160^{+750}_{-520}$\\
170403+23:06:11UTC\ias           &  1175295989.23           &        0.03           &         .07           &          HL           &        7.26           &         5.2           &         5.2           &           -           & $53^{+23}_{-13}$ & $35^{+13}_{-15}$ & $-0.20^{+0.35}_{-0.37}$ & $2500^{+2100}_{-1300}$\\
170629+04:13:55UTC           &  1182744853.11           &        0.02           &         .06           &          HL           &        6.72           &         6.6           &         4.8           &           -           & $49^{+20}_{-30}$ & $7.3^{+4.6}_{-2.6}$ & $0.73^{+0.15}_{-0.98}$ & $1880^{+1450}_{-940}$\\
170620+01:14:02UTC           &  1181956460.10           &        0.02           &         .04           &          HL           &        6.18           &         5.7           &         5.1           &           -           & $29.4^{+13.2}_{-6.8}$ & $17.9^{+5.4}_{-5.5}$ & $0.05^{+0.25}_{-0.25}$ & $1710^{+1300}_{-850}$\\
170801+23:28:19UTC           &  1185665317.35           &           -           &         .04           &          LV           &        8.59           &           -           &         6.9           &         4.3           & $23.9^{+12.6}_{-6.6}$ & $12.4^{+4.7}_{-4}$ & $-0.09^{+0.25}_{-0.24}$ & $1070^{+920}_{-580}$\\
170818+09:34:45UTC\footnote{Parameter estimates for this candidate are derived only from the LIGO-Hanford and Virgo detectors. LIGO-Livingston was operating at the time, but did not produce a trigger that contributed to the event (see discussion in Sec.(\ref{sec:bbh}))}           &  1187084103.28           &           -           &         .04           &          HV           &        8.40           &         6.5           &           -           &         4.4           & $55^{+59}_{-28}$ & $23^{+43}_{-15}$ & $0.06^{+0.48}_{-0.45}$ & $3100^{+1700}_{-1900}$\\
\end{tabular}
}
  \end{center}
\footnotesize{\label{prev2}$^x$Also identified in GWTC-1~\citep{LIGOScientific:2018mvr}, $^y$1-OGC~\citep{Nitz:2018imz}, or $^z$\cite{Venumadhav:2019tad,Venumadhav:2019lyq}}
\end{table*}

\section{Observational Results}

We present compact binary merger candidates from the complete set of public LIGO and
Virgo data spanning the observing runs from 2015-2017. This comprises roughly 171 days of
multi-detector observing time which we divide into 31 sub-analyses. Except as noted,
each analysis contains $\sim5$ days of observing time which allows for estimation of the
false alarm rate to $<$ 1 per 10,000 years. This interval allows us to track changes in the detector configuration which may result in time-changing detector quality. All data was retrieved from GWOSC~\cite{Vallisneri:2014vxa}, and we have used the most up-to-date version of bulk
data released. We note that an exceptional data release was produced by GWOSC which contains background data relating to GW170608. We have analyzed this data release separately to preserve consistent data quality.

The top candidates sorted by FAR from the complete analysis are given in Table~\ref{table:complete}. All of the most significant candidates were observed by LIGO-Hanford and LIGO-Livingston which are the two most sensitive detectors in the network and contribute the bulk of the observing time. There are 8 BBH and 1 BNS candidates at a FAR less than 1 per 100 years. These sources are confidently detected in the full analysis without optimizing the search for any specific population of sources. The most significant following candidates correspond to GW170729, GW170121, GW170727 and GW170818 respectively. A similar PyCBC-based analysis was performed in~\cite{LIGOScientific:2018mvr} but used a higher single-detector SNR threshold than employed in our analysis ($\rho > 5.5$ vs 4.0); as the latter three events were found with $\rho <= 5.1$ in the LIGO-Hanford detector, we would not expect this earlier analysis to identify them.
\begin{table*}
    \caption{Candidate events with template parameters consistent with BNS mergers sorted by ranking statistic \rankingstat. The chirp mass $\mathcal{M}$ of the candidate's associated template waveform is given in the detector frame. All candidates here were found by the LIGO-Hanford and LIGO-Livingston observatories. The table lists the false alarm rate for each candidate in the context of the full search (FAR$_{FULL}$) or just the selected BNS region (FAR$_{BNS}$).}
    \label{table:bns}
\begin{center}
\begin{tabular}{cllllrrr}
Date Designation & GPS Time & \rankingstat{} &  FAR$_{FULL}^{-1}$ (y) &  FAR$_{BNS}^{-1}$ (y) & $\rho_H$ & $\rho_L$ & $\mathcal{M}$ \\ \hline
170817+12:41:04UTC          & 1187008882.45          &     180.46          &   $>10000$          &   $>10000$          &       18.6          &       24.3          &       1.20         \\
161217+02:44:33UTC          & 1165977890.44          &      10.81          &        .01          &        .27          &        6.2          &        6.0          &       1.15         \\
151214+21:03:35UTC          & 1134162232.89          &       9.26          &       .003          &        .12          &        6.0          &        5.7          &       1.06         \\
151105+20:34:28UTC          & 1130790885.49          &       8.97          &       .002          &        .03          &        6.0          &        6.3          &       1.29         \\
160103+02:29:54UTC          & 1135823411.78          &       8.41          &       .002          &        .04          &        5.3          &        6.5          &       1.16         \\
170204+00:34:28UTC          & 1170203686.48          &       8.40          &       .002          &        .06          &        5.1          &        6.4          &       1.25         \\
170819+11:06:26UTC          & 1187176004.83          &       8.37          &       .004          &        .12          &        6.2          &        6.0          &       1.06         \\
170213+21:45:15UTC          & 1171057533.29          &       8.35          &       .002          &        .06          &        7.4          &        6.1          &       1.15         \\
151112+05:48:49UTC          & 1131342546.36          &       8.11          &      .0008          &        .01          &        5.7          &        6.4          &       1.35         \\
150930+12:45:03UTC          & 1127652320.31          &       8.10          &       .001          &        .04          &        6.0          &        5.8          &       1.15         \\
\end{tabular}
\end{center}
\end{table*}

\subsection{Binary Black Holes}
\label{sec:bbh}
\begin{figure*}
    \centering
    \includegraphics[width=0.49\textwidth]{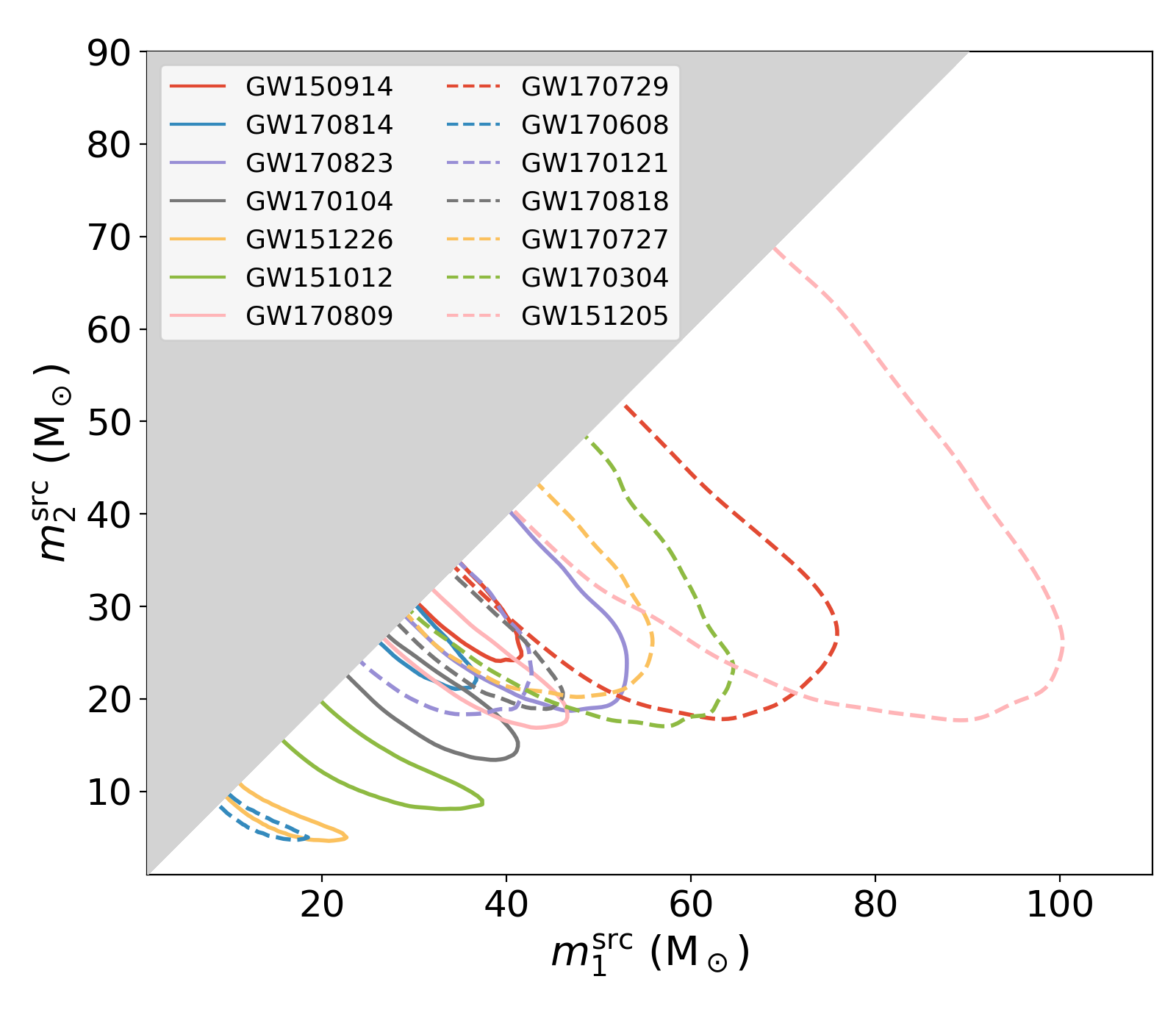}
    \includegraphics[width=0.49\textwidth]{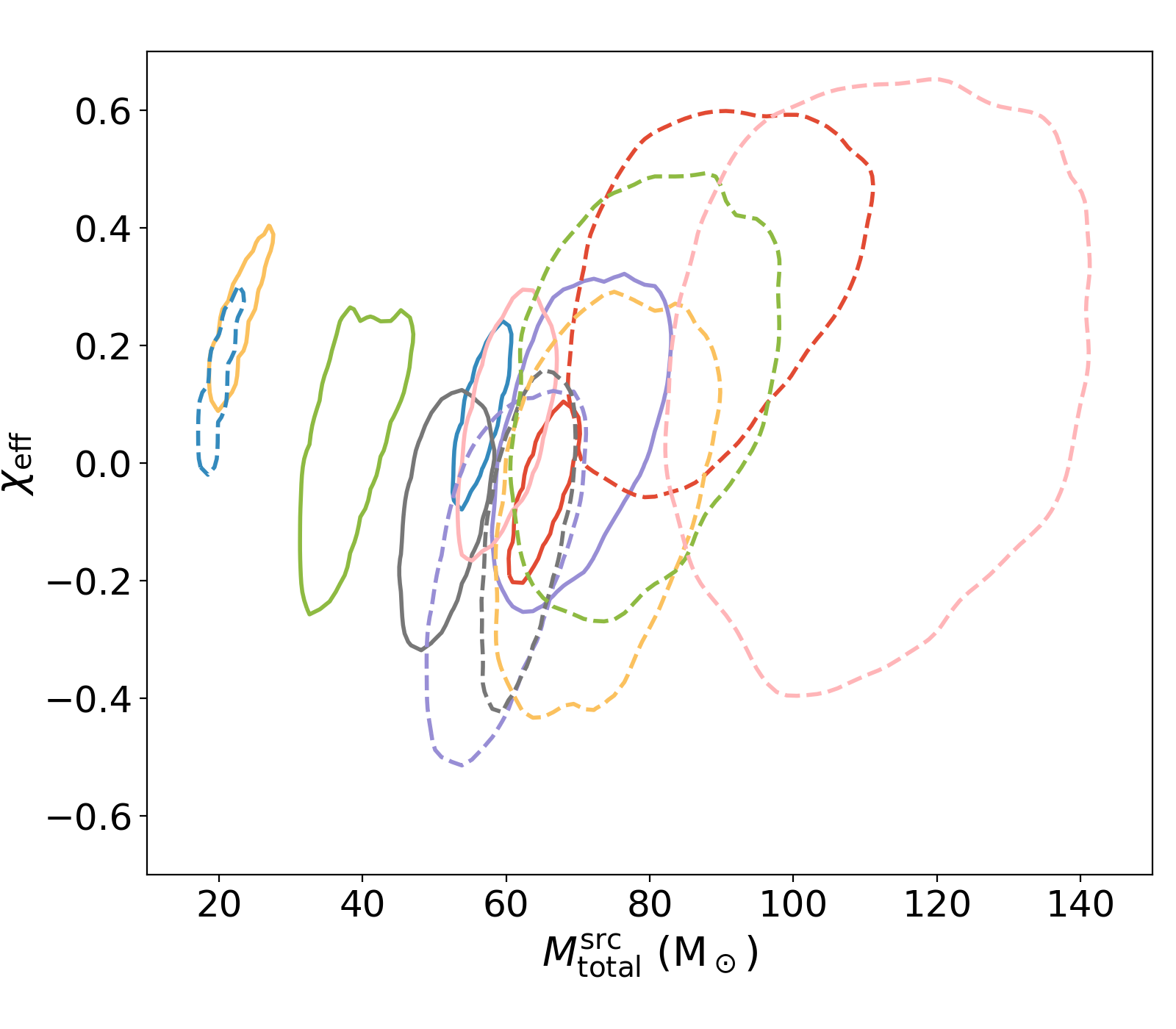}
    \caption{
    Marginalized $90\%$ credible region for all binary black hole candidates with $\pastro \geq 0.5$ in source-frame component masses (left) along with source-frame total mass and effective spin (right). GW170121, GW170304 and GW170727 which were previously reported in~\cite{Venumadhav:2019lyq} are broadly consistent with the existing population of observed BBH mergers. GW151205, a new BBH candidate with $\pastro\sim0.53$, is likely the most massive merger reported to date if astrophysical.}
    \label{fig:bbh_posteriors}
\end{figure*}

\begin{figure}
    \centering
    \includegraphics[width=\columnwidth]{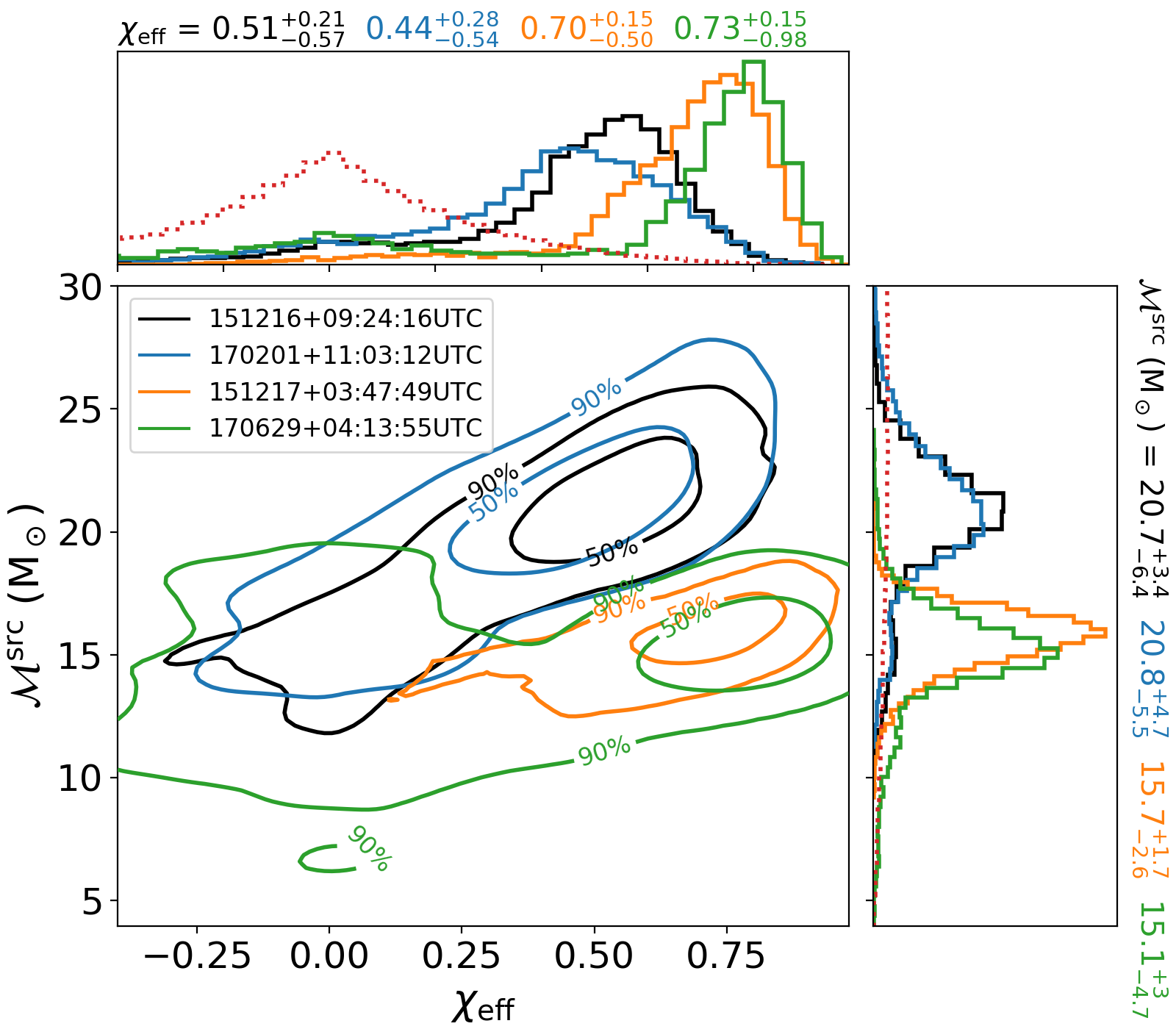}
    \caption{Comparison of marginalized $\chieff$ and source-frame chirp mass $\mathcal{M}^{\rm src}$ posteriors between 151216+09:24:16UTC, 170201+11:03:12UTC, 151217+03:47:49UTC, and 170629+04:13:55UTC. The marginalized prior on each parameter is shown by the dotted red lines. Contrary to \cite{Zackay:2019tzo}, we find that 151216+09:24:16UTC has support at zero effective spin. The candidate bears striking resemblance to 170201+11:03:12UTC, and, to a lesser extent, 151217+03:47:49UTC and 170629+04:13:55UTC. All four have $\chieff$ posteriors that diverge strongly from the prior (with peaks between $\sim0.5$ and $\sim0.7$) and similar chirp masses, which distinguishes them from the other BBH candidates in Table~\ref{table:bbh}. This may indicate a new population of binary black holes, or a common noise feature. If the former, ongoing and future observing runs should yield more candidates with similar properties and larger astrophysical significance.}
    \label{fig:151216_151217_170201_comparison}
\end{figure}

Using the targeted BBH analysis introduced in Sec.~\ref{sec:bbhsearch} we report
results for BBH mergers consistent with the existing set of highly significant merger events in Table~\ref{table:bbh}. The probability that a candidate is astrophysical in origin, \pastro, is calculated for the most significant candidates. Our analysis identifies 14 BBH candidates with $\pastro > 50\%$, meeting the standard detection criteria introduced in~\cite{LIGOScientific:2018mvr} and similarly followed in~\cite{Venumadhav:2019lyq}. Our results are broadly consistent with the union of those two analyses as our candidate list includes all previously claimed BBH detections.
We confirm the observation of GW170121, GW170304, and GW170727 reported in~\cite{Venumadhav:2019tad} as significant. We also report the marginal detection of GW151205.

Several marginal events reported in~\cite{Venumadhav:2019lyq,Venumadhav:2019tad} are found as top candidates, but do not meet our detection threshold based on estimated probability of astrophysical origin. Numerous differences between these two analyses --- including template bank placement, treatment of data, choice of signal consistency test, and method for assigning astrophysical significance --- may be the cause of reported differences. The consistency of results for less marginal candidates indicates that differences in analysis sensitivity are likely marginal. Cross comparison with a common set of simulated signals would be required for a more precise assessment.

Future analyses incorporating more sophisticated treatment of the source distribution may yield different results for the probability of astrophysical origin for some sub-threshold candidates. For example 151216+09:24:16UTC, which was first identified in~\cite{Nitz:2018imz} and is now assigned a $\pastro\sim0.2$, could obtain a higher probability of being astrophysical under a model with a distribution of detected mergers peaked close to its apparent component masses, rather than uniform over $\mathcal{M}$ as taken here.
In any case the astrophysical probability we assign assumes that the candidate event, if astrophysical, is drawn from an existing population. The prior applied here to the population distribution over component masses could be extended to the distribution over component-object spins. (Here, we implicitly apply a prior over spins which mirrors the density of templates, which is not far from uniform over $\chieff$.) As 151216+09:24:16UTC may have high component spins, if the set of highly significant observations does not include any comparable systems its probability of astrophysical origin could be arbitrarily small, depending on a choice of prior distribution over spins.

We infer the properties of our BBH candidates using Bayesian parameter inference implemented by the PyCBC library~\citep{biwer:2018osg}. We use the IMRPhenomPv2 model which describes the dominant gravitational-wave mode of the inspiral-merger-ringdown of precessing non-eccentric binaries~\citep{Schmidt:2014iyl, Hannam:2013oca}. For each candidate, we use a prior isotropic in sky location and binary orientation. As in \cite{LIGOScientific:2018mvr}, our prior on each component object's spin is uniform in magnitude and isotropic in orientation.

Since many of the candidates are at large ($>1\,$Gpc) distances, we assume a prior which is uniform in comoving volume, and a prior uniform in \emph{source-frame} component mass. We use standard $\Lambda$CDM cosmology~\citep{Ade:2015xua} to relate the comoving volume to luminosity distance, and to redshift the masses to the detectors' frame. This choice of prior differs from previous analyses~\citep{LIGOScientific:2018mvr,Venumadhav:2019lyq,Venumadhav:2019tad}, which used a prior uniform in volume (ignoring cosmological effects) and detector-frame masses. A prior uniform in co-moving volume assigns lower weight to large luminosity distances than a prior uniform in volume. Consequently, the luminosity distances we obtain for some candidates is slightly lower than previously reported values (e.g., we obtain $D_L = 2300^{+1600}_{-1200}\,$Mpc for GW170729, whereas \cite{LIGOScientific:2018mvr} obtained $D_L = 2840^{+1400}_{-1360}\,$Mpc).

The marginalized parameter estimates of the component masses, effective spin, and luminosity distance for the top 30 BBH candidates are given in Table~\ref{table:bbh}. Plots of the marginalized posteriors for the BBH candidates with $\pastro \geq 0.5$ is show in Fig.~\ref{fig:bbh_posteriors}. For candidates previously reported by the LVC, our results broadly agree with existing parameter estimates~\citep{LIGOScientific:2018mvr,De:2018zrk}. Similarly, we find no clear evidence for precession in our candidates. \cite{Venumadhav:2019lyq} and ~\cite{Zackay:2019tzo} reported marginal high-mass BBH candidates, in particular 170403+23:06:11UTC with $\chieff=-0.7^{+0.5}_{-0.3}$ and 151216+09:24:16UTC with $\chieff=0.8^{+.15}_{-.21}$, which excludes $\chieff\sim0$.  In addition to assigning these candidates lower astrophysical significance, we find that $\chieff\sim0$ is excluded for neither candidate. For 151216+09:24:16UTC, we find several points in the posterior around $\chieff\sim 0$ with likelihood values similar to that around $\chieff\sim 0.5$. This indicates that the discrepancy in $\chieff$ 
between our analysis and that of \cite{Venumadhav:2019lyq} cannot be entirely explained by differences in prior choice; the difference may be due to differing analysis methods.

We find three other events with $\pastro < 0.3$ that have $\chieff$ and masses similar to that of 151216+09:24:16UTC. These are illustrated in Fig.~\ref{fig:151216_151217_170201_comparison}. The four events differ from the other events listed in Table~\ref{table:bbh} in that the posterior distribution of $\chieff$ strongly deviates from the prior, with the peak in the posterior between $\chieff \sim 0.5$ and $\sim0.7$. All four events also have similar chirp masses. If these events are from a new population of binary black holes, then ongoing and future observing runs should yield candidates with similar properties at high astrophysical significance. Alternatively, they may indicate a common noise feature selected by our analysis.

GW151205, a BBH merger with $\pastro \sim 0.53$, may challenge standard stellar formation scenarios if astrophysical. Models that account for pulsational pair instability supernovae (PPISNe) or pair-instability supernovae (PISNe) in stellar evolution suggest the maximum mass of the remnant black hole is $\sim40-50\,\msun$~\citep{Woosley:2016hmi,Belczynski:2016jno,Marchant:2018kun,Woosley_2019,Stevenson:2019rcw}. We estimate that there is $>95\%$ probability that the primary black hole has a source-frame mass $>50\,\msun$, which may suggest formation through an alternate channel such as hierarchical merger. Studies have proposed that GW170729 may have a similar origin~\citep{Khan:2019kot, Yang:2019cbr, Kimball:2019mfs}. However, \cite{Fishbach:2019ckx} showed that when all of the BBHs are analyzed together, GW170729 is consistent with a single population of binaries formed from the standard stellar formation channel. Likewise, GW151205 will need to be analyzed jointly with the other events to determine if there are one or more populations present.

The least significant candidate in the targeted BBH analysis, 170818+09:34:45UTC, was identified in the LIGO-Hanford and Virgo detectors by the search pipeline; the parameter estimates in Table~\ref{table:bbh} are derived using these observatories alone. However, the LIGO-Livingston detector was operational at the time of the event. Our search does not currently enforce that a candidate observed only in a subset of detectors is consistent with lack of observation in the others. We find that if LIGO-Livingston is included in the parameter estimation analysis, the log likelihood ratio is significantly reduced. This suggests that the event is not astrophysical in origin.

\subsection{Neutron Star Binaries}
Our analysis identified GW170817 as a highly significant merger, however, no further individually significant BNS nor NSBH mergers were identified. As the population of BNS and NSBH sources is not yet well constrained, we cannot reliably employ the methodology used to optimize search sensitivity to an astrophysical BBH merger distribution.
However, BNS candidates especially are prime candidates for the observation of electromagnetic counterparts such as GRBs and kilonovae. It may be possible by correlating with auxiliary datasets to determine if weak candidates are astrophysical in origin. An example is the sub-threshold search of Fermi-GBM and 1-OGC triggers~\citep{Nitz:2019bxt}, which defined, based on galactic neutron star observations~\citep{Ozel:2012ax}, a likely BNS merger region to span $1.03 < \mathcal{M} < 1.36$ and effective spin $|\chieff| < 0.2$. This region is highlighted in Fig.~\ref{fig:bank} and the top candidates are shown in Table~\ref{table:bns}.

\label{sec:nsobs}
\section{Data Release}
We provide supplementary materials online which provide information on each of $\sim10^6$ sub-threshold candidates~\citep{2-OGC}. Reported information includes candidate event time, SNR in each observatory, and results of the signal-consistency tests performed. A separate listing of candidates within the BBH region discussed in Sec.~\ref{sec:bbhsearch} is also provided, including estimates of the probability of astrophysical origin \pastro\ for the most significant of these candidates. To help distinguish between these large number of candidates, our ranking statistic and estimate of the false alarm rate are also provided for every event. Configuration files for the analyses performed and analysis metadata are also provided. For the 30 most significant BBH candidates, we also release the posterior samples from our Bayesian parameter inference.

\section{Conclusions}
The 2-OGC catalog of gravitational-wave candidates from compact-binary coalescences spanning the full range of binary neutron star, neutron star--black hole, and binary black hole mergers is an analysis of the complete set of LIGO and Virgo public data from the observing runs in 2015-2017. A third observing run (O3) began in April, 2019~\cite{Aasi:2013wya}. Alerts for several dozen merger candidates have been issued to date during this  run\footnote{https://gracedb.ligo.org/superevents/public/O3}. The first half of the run (O3a) ended on Oct~1 2019 with a planned release of the corresponding data in Spring 2021. As the data is not yet released, the catalog here covers only the first two observing runs.

We use a matched-filtering, template-based approach to identify candidates and improve over the 1-OGC analysis~\citep{Nitz:2018imz} by incorporating corrections for time variations in power spectral density estimates and network sensitivity. Furthermore, we have demonstrated extending a PyCBC-based analysis to handle data from more than two detectors. The 2-OGC catalog contains the most comprehensive set of merger candidates to date, including 14 BBH mergers with $\pastro > 50\%$ along with the single BNS merger GW170817. We independently confirm many of the results of~\cite{LIGOScientific:2018mvr} and ~\cite{Venumadhav:2019lyq}. We find no additional individually significant BNS or NSBH mergers, however, we provide our full set of sub-threshold candidates for further analysis\citep{2-OGC}.

\acknowledgments
We acknowledge the Max Planck Gesellschaft and the Atlas cluster computing team at AEI Hannover for support. Research supported by Maria de Maeztu Unit of Excellence MDM-2016-0692. This research was supported in part by the National Science Foundation under Grant No. NSF PHY-1748958.
This research has made use of data, software and/or web tools obtained from the Gravitational Wave Open Science Center (https://www.gw-openscience.org), a service of LIGO Laboratory, the LIGO Scientific Collaboration and the Virgo Collaboration. LIGO is funded by the U.S. National Science Foundation. Virgo is funded by the French Centre National de Recherche Scientifique (CNRS), the Italian Istituto Nazionale della Fisica Nucleare (INFN) and the Dutch Nikhef, with contributions by Polish and Hungarian institutes.
\bibliography{references}

\end{document}